# Yield forecasting with machine learning and small data: what gains for grains?

Michele Meroni[a*1], François Waldner[a1], Lorenzo Seguini[a], Hervé Kerdiles[a], Felix Rembold[a]

[a] European Commission, Joint Research Centre (JRC), Via E. Fermi 2749, I-21027 Ispra (VA), Italy.

* Michele.meroni@ext.ec.europa.eu

[1] These authors contributed equally to this work.

**ABSTRACT**

Forecasting crop yields is important for food security, in particular to predict where crop production is likely to drop. Climate records and remotely-sensed data have become instrumental sources of data for crop yield forecasting systems. Similarly, machine learning methods are increasingly used to process big Earth observation data. However, access to data necessary to train such algorithms is often limited in food-insecure countries. Here, we evaluate the performance of machine learning algorithms and small data to forecast yield on a monthly basis between the start and the end of the growing season. To do so, we developed a robust and automated machine-learning pipeline which selects the best features and model for prediction. Taking Algeria as case study, we predicted national yields for barley, soft wheat and durum wheat with an accuracy of 0.16-0.2 t/ha (13-14 % of mean yield) within the season. The best machine-learning models always outperformed simple benchmark models. This was confirmed in low-yielding years, which is particularly relevant for early warning. Nonetheless, the differences in accuracy between machine learning and benchmark models were not always of practical significance. Besides, the benchmark models outperformed up to 60% of the machine learning models that were tested, which stresses the importance of proper model calibration and selection. For crop yield forecasting, like for many application domains, machine learning has delivered significant improvement in predictive power. Nonetheless, superiority over simple benchmarks is often fully achieved after extensive calibration, especially when dealing with small data.

**Keywords:** Remote sensing, agriculture, yield forecasting, meteorological data, machine learning.

## 1. Introduction

Timely and reliable crop yield forecasts play an important role in supporting national and international agricultural and food security policies, stabilizing markets and planning food security interventions in food-insecure countries (Becker-Reshef et al., 2020). Monitoring crop status, growth and productivity helps establishing food emergency responses and planning for a long-term, sustainable development strategy. In particular, with increasing impact weather variability and extremes on food security (FAO et al., 2018), governments need to anticipate crop production losses to respond appropriately.

Operational yield forecasting approaches are often based on empirical regression models linking historical yields and administrative units-averages of seasonal satellite and climate data for cultivated areas (Schauberger et al., 2020). In operations, the model is then fed with data observed for the current growing season to forecast the final yield. Satellite instruments providing frequent, coarse resolution satellite image time series, such as AVHRR (Advanced Very High Resolution Radiometer), SPOT-VGT (SPOT-VEGETATION), or MODIS (Moderate Resolution Imaging Spectroradiometer), have been extensively used for yield estimation at regional scales (Atzberger et al., 2016; Rembold et al., 2013). Typically, yields are estimated by regressing either vegetation indices or crop biophysical variables at specific dates, which are proxies for green biomass, or features characterising the dynamics of a vegetation index over time such as the senescence or the green-up rate (Waldner et al., 2019). Popular linear regression approaches use the Normalised Difference Vegetation Index (NDVI; Rouse et al., 1974) either at its peak (*e.g.*, Becker-Reshef et al., 2010; Franch et al., 2015) or its cumulative value over the growing season (*e.g.*, López-lozano et al., 2015; Meroni et al., 2013).

Whereas linear regressions may fail to capture the complex interactions between environmental conditions and yield, machine learning (ML) models have demonstrated powerful performance in various data-driven applications, including yield estimation (Cai et al., 2019; Johnson et al., 2016; Kamir et al., 2020; Mateo-Sanchis et al., 2019; Wolanin et al., 2020; Zhang et al., 2020). A large set of algorithms is now available for regression (*e.g.*, random forest, support vector regression, kernel machines and neural networks). Potentially, also deep learning methods employing multiple layers of computation in neural networks (Goodfellow et al., 2016) could be used to exploit heterogeneous information (*i.e.*, remote sensing and meteorological data) and find complex, nonlinear relations with crop yield. Machine learning and deep learning are data-driven. They typically need large data sets to provide accurate forecasts across a range of conditions. However, regional crop yield forecasting currently operates in data-poor context. Labels (*i.e.*, yield data) are available for a small number of administrative regions and the amount archive satellite data depends on the sensor of interest (*e.g.*, 20 years for MODIS). Increased spatial granularity would not necessarily proportionally increase the information content because large yield variations are modulated by weather phenomena that are typically covariate in space (*e.g.*, droughts). While the small size of yield data sets limits the use of deep learning, it remains challenging for ML models. Therefore, as sample size is critical to achieve accurate and reliable predictions with ML models, potential advantages of using ML instead of simpler linear regression models needs to be verified empirically.

The use of ML models in operational settings poses also practical difficulties related to the set-up of the modelling framework. Typical ML workflows include model section, feature engineering and selection, hyper-parameter optimisation, and model testing in a way that is relevant for the application. Operating in data-poor contexts complicates the latter task as avoiding information leakage between the training and test sets may restrict the available options.

In addition to such ML specific considerations, a barrier in the test and operational application of such methods is due to the fact that the interested practitioner is upfront confronted with the problem of selecting the appropriate explanatory variables among the plethora of existing remote sensing and meteorological data sources. After that, data must be downloaded and a processing chain must be set up to extract the information at the relevant administrative unit level from the typically gridded raster data.

This work proposes an end-to-end modelling framework to facilitate the test and application of ML models for operational yield forecasting at the administrative level. It is based on the specific requirements of the application and can be transferred to any country. The framework capitalises on public, global satellite and climate data provided in a ready-to-use format by the Anomaly hotSpot of Agricultural Production (ASAP) early warning decision support system of the European Commission Joint Research Centre (Meroni et al., 2019b; Rembold et al., 2018). Our modelling framework automatically defines the temporal domain of the forecast operations using land surface phenology, tests a suite of state-of-art ML algorithms in a number of configurations, produces accuracy figures tailored to operational yield forecasting and compare them against simple benchmark models. As a case study, we deployed the modelling framework in hindcast (2002-2018) to predict the yield of Algeria's main cereals (durum and soft wheat and barley).

## 2. Study area and data

### 2.1. Study area

Our study area encompasses 20+ provinces (wilayas) in Algeria. The main cereal crops are durum wheat (average national production = 1.8 Mt), barley (1.2 Mt) and soft wheat (0.75 Mt). These cereals are generally rainfed (only 3% of the cereal area is irrigated according to FAOSTAT). Sowing takes place in autumn (between October and November depending on autumn rainfall); harvests occur in spring to early summer (from May to July) and tend to begin earlier in the south. Yields generally range from 0.5 t/ha to 2.5 t/ha (Fig. 1 and S1) and are highly influenced by climatic inter-annual variability (Benmehaia et al., 2020). Climate ranges from desert in the South, to Mediterranean in the North, with also an east-west gradient with drier areas in the west. Average yearly rainfall (2001-2018) ranges from 106 mm in the South (province of Biskra) to 500 mm in the North (province of Constantine). The set of provinces selected varies by crop type. For each crop, we ranked all the provinces by decreasing average production and retained those that contributed to the 90% of the national mean crop production (Fig. S2), thus excluding marginal cropping provinces. The number of provinces covered is 24, 23 and 20 for durum wheat, barley and soft wheat, respectively (Table S1).

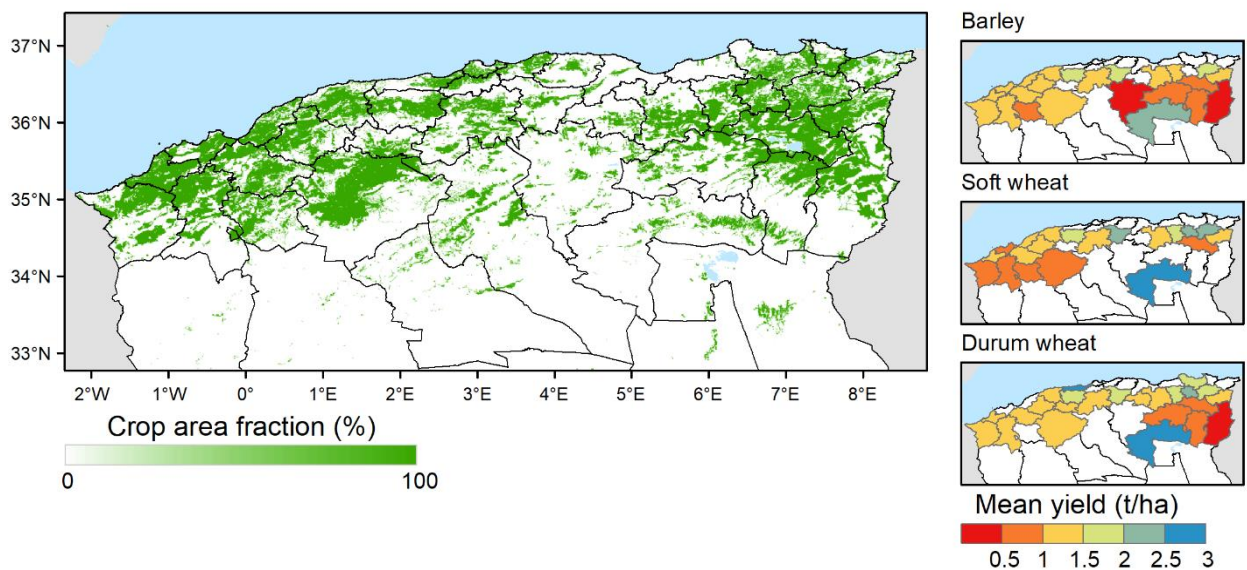

**Fig. 1.** Location of the study area in Algeria. Left: percent crop area fraction. Right: provinces covered by crop and their mean yield.

### 2.2. Yield data

Official yield statistics for the three crops were provided by the Direction des Statistiques Agricoles et des Systèmes d'Information - Ministère de l'Agriculture et du Développement Rural (DSASI-MADR). Yield statistics are obtained from ground sample surveys and are available as administrative-level averages of grain weight per unit of sowed area. The analysis was performed over the period 2002-2018. No statistically relevant yield trend was detected during the analysed period at national level and for most of the provinces (yield time series and statistical tests in Supplementary Material Fig. S3). The time range of the analysis was determined by the availability of both MODIS imagery and wheat yield official statistics. As a result, the total number of yield data points ($n$ = 17 years × no. provinces) is 408, 391 and 340 for durum wheat, barley and soft wheat, respectively.

### 2.3. Remotely-sensed and climate data

We estimated province level crop yield using as input a set of time series of satellite-derived vegetation and meteorological variables freely downloadable from the ASAP early warning system (https://mars.jrc.ec.europa.eu/asap/download.php; Table 1), which conveniently already serves the data as crop area-averaged temporal time series by administrative units. The data set includes NDVI as a proxy of green biomass and the main meteorological variables influencing crop growth: precipitation, temperature and global radiation. ASAP data are derived from various standard sources (MODIS for satellite observation of vegetation status, CHIRPS and ECMWF for meteorological data), tailored to crop monitoring applications and comparable to those of other monitoring systems (Becker-Reshef et al., 2020; Fritz et al., 2019). Variables were downloaded from the ASAP system at GAUL (Global Administrative Units Layers of the Food and Agriculture Organization) level 1, which in Algeria corresponds to provinces (wilayas). Pixel level values of input variables are spatially aggregated to the province level as the weighted average according to fractional area occupied by cropland in each pixel.

**Table 1.** Input data used in this study. Dekads refers to the nearly 10-day periods spanning day 1-10, day 11-20, day 21-end of the month.

| Category | Variable | Spatial resolution | Temporal resolution | Source |
|---|---|---|---|---|
| Remote sensing of vegetation | NDVI | 1 km | dekad | MODIS MOD13A2 and MYD13A2 V006 filtered using the weighted Whittaker smoother as described in Meroni et al. (2019a) |
| Meteorological data | Average air temperature (at | 25 km | dekad | Elaboration on ECMWF ERA5 data as described in Meroni et al. (2019b) |
|  | 2 m), Global Radiation sum |  |  |  |
|  | Precipitation sum | 5 km | dekad | CHIRPS 2.0 (Funk et al., 2015) |
| Cropland fraction | Percentage of the pixel occupied by cropland | 1 km | Static | Derived from hybrid cropland mask combining multiple land cover maps (Pérez-Hoyos et al., 2020) |

### 2.4. Time domain and temporal aggregation

Our analysis was performed over the average growing season period. This period was determined using satellite-derived land surface phenology derived from the NDVI time series using a model fit approach (Meroni et al., 2014; Zhang et al., 2003). Here we used a double logistic function to fit the NDVI trajectory and derive the following metrics: start of season (the time when fitted NDVI rises above 20% of the ascending amplitude of the seasonal profile) and end of the season (when fitted NDVI drops below 20%). These metrics were extracted for all the investigated provinces to obtain an average start and end of season (the start of season was the second dekad of November ± 2 dekads; the end of season was second dekad of June ± 1 dekad). We thus aggregated input data by month from November to June and computed their average, maximum, minimum or sum (Table 2). We then normalised these features using standard scores prior to their ingestion in the machine learning workflow.

**Table 2.** Definition of operators used in monthly aggregation and resulting input features.

| Category | Variable | Aggregation | Feature name |
|---|---|---|---|
| Satellite observation of vegetation | NDVI | Average, maximum | ND, $ND_{max}$ |
| Meteorology | Temperature | Average, maximum, minimum | T, $T_{min}$, $T_{max}$ |
|  | Global Radiation | Sum | Rad |
|  | Precipitation | Sum | Rain |

## 3. Methods

Yield forecasts are computed at the beginning of each month from December to July, when data of the last 10-day period of the previous month becomes available. Forecasts use features generated from the start of the season until the month preceding the forecast (included), so that for instance the April forecast used data collected until end of March. There are thus eight forecasts per season. For testing purposes, forecasts at month $m$ are made in hindcasting, *i.e.*, using monthly remote sensing and meteorological data from November up to month $m-1$ included. A nested cross-validation is used to assess forecasting performances avoiding information leak (Section 3.1.1).

Several ML model configurations were tested. A model configuration is here defined by a set of input features and feature selection options (Section 3.1.2) and a ML algorithm (Section 3.1.3). In parallel, two benchmark models are employed: the peak NDVI and the null model (Section 3.2). After that, the best performing ML model configurations are selected and compared with benchmark models at each forecasting time (Section 3.3.1). The practical significance of the performance difference among the ML and benchmark models is then tested (Section 3.3.2). Finally, we investigate the effect of adding categorical variables capturing province-specific characteristics and feature selection on model predictive performances (Section 3.3.3).

### 3.1. Machine learning workflow

#### 3.1.1. Cross-validation strategy

Our experimental context can be defined as data poor ($n$ = 408, 391 and 340 province-year data points for durum wheat, barley and soft wheat). In such conditions, cross-validation is essential to estimate model accuracy while making optimal use of the data available. ML models set-up requires to determine both the hyper-parameters and the coefficients of the model. Thus, three independent datasets are required: the training set used to train the model, the validation set used to optimise the set of hyper-parameters, and the test set used to estimate the performance of the optimised model in prediction.

Cross-validation has to be designed bearing in mind the scope of the application, here forecasting the yield of a season that has not been previously seen by the model. Randomly splitting the available data set is thus not a viable option for cross-validation because all years would be present in all sets, resulting in information leakage from the testing set to the training one.

In such settings, performance assessment requires splitting the data twice in a nested leave-one-year-out cross-validation (Fig. 2). In the outer cross-validation loop, data of one year of the $n$ available is held out. The remaining $n$-$1$ years are subjected to an inner cross-validation loop. In this inner-loop, data of one year of the $n$-$1$ available is held out at a time for validation. Model hyper-parameters are selected within the inner cross-validation loop, thus using $n$-$2$ years to estimate the held out at each iteration. After that, all $n$-$1$ years are used to determine model coefficients (*i.e.*, model training) and then the obtained model is used to predict the yield of the outer-loop held out year (*i.e.*, model testing). This procedure is repeated for all the $n$ years to assess the performance of the model in prediction.

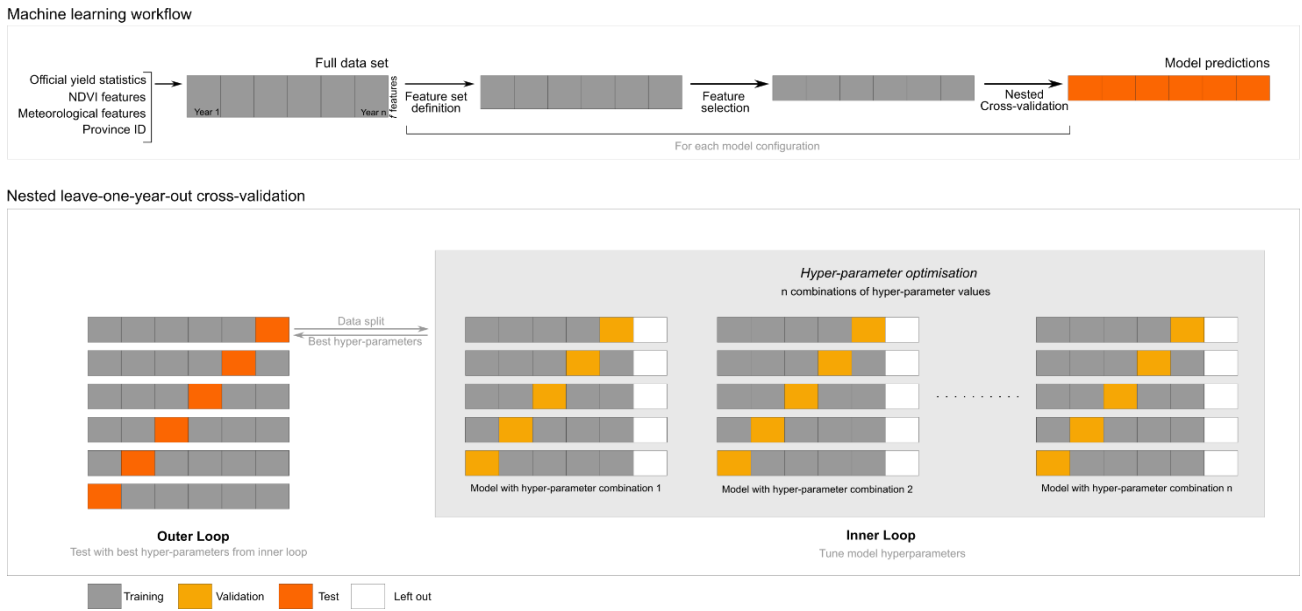

**Fig. 2.** Nested leave-one-year-out cross-validation loop used in the ML workflow. Top panel: overall workflow for testing model configurations. Bottom panel: representation of the nested cross-validation loop for a single model configuration.

Following this approach, we trained $n$ different models: one for each of the $n$ outer loop year. Thus, in each iteration of the outer-loop different hyper-parameters and model coefficients can be selected. In operations, however, we would use the $n$ available year to train the model after having validated the model hyper-parameters with a single leave-one-full-year out cross-validation.

#### 3.1.2. Model inputs

Feature selection is an important step in machine learning workflows. It can improve prediction accuracy by discarding irrelevant features, accelerate model training and inference and reduce maintenance of the feature ingestion pipeline. We tested a range of model inputs by manually defining input feature sets, automatically selecting features and adding categorical variables capturing province-specific characteristics.

Two successive stages of feature selection are tested in the modelling framework. First, from the full set of available remote sensing and meteorological explanatory variables (named RS&Met) we define some relevant subsets (Table 3). To explore the contribution of satellite and meteorological data we isolated the respective variables in the feature sets RS and Met. After that, we defined three additional sets (RS&Met-, RS- and Met-) by retaining the most essential variables for each of the original set.

**Table 3.** Variables considered in the manually engineered feature sets.

| Set name | Variables used | | | | | | |
|---|---|---|---|---|---|---|---|
| | Remote sensing | | Metereology | | | | |
| | NDVI (avg) | NDVI (max) | Rad (sum) | Rain (sum) | T (avg) | T (min) | T (max) |
| *RS&Met* | • | • | • | • | • | • | • |
| *RS* | • | • | | | | | |
| *Met* | | | • | • | • | • | • |
| *RS&Met-* | • | | | • | • | | |
| *RS-* | • | | | | | | |
| *Met-* | | | | • | • | • | |

Second, we implemented an additional feature selection step. We chose the Minimum Redundancy and Maximum Relevance approach (mRMR; Peng et al., 2005), which selects relevant features while controlling for their redundancy. As a filter method, mRMR is fast because it does not involve any model training. The percentage of features to be retained was treated as a hyper-parameter of the model, *i.e.*, it was optimised in the inner cross-validation loop using an exhaustive search on the following values: 5, 10, 25, 50, 75, and 100 % of input features.

Remote sensing and meteorological features do not contain information about soil, management practices, and other unobserved variables that can influence the relation between features and yield at the province level. One way to convey this province-specific information to the model is adding a categorical variable representing the province. The variable is one-hot encoded (*i.e.*, replaced by new binary variables, one per unique categorical value), resulting in 24, 23 and 20 additional one-hot encoded administrative unit binary features (OHEau) for durum wheat, barley and soft wheat, respectively.

In summary, at each forecast month, all ML algorithms are trained on six input feature sets, with and without mRMR feature selection (6 fractions of input features tested) and with and without OHEau, for a total of 84 different combinations tested.

### 3.1.3. Machine learning algorithms

We selected and compared five standard machine learning regression algorithms: least absolute shrinkage and selection operator (LASSO; Tibshirani, 1996), Random Forest (RF; Breiman, 2001), multi-layer perceptron (MLP; Rosenblatt, 1961), support vector regression with linear and radial basis function kernels (SVR lin and SVR rbf; Vapnik et al., 1997), and gradient boosting (GBR; Friedman, 2001). Algorithms and pipelines were implemented using the python package scikit-learn (Pedregosa et al., 2011). Each model has a set of hyper-parameters. Some were optimised using an exhaustive search from a grid of empirical candidates (Table 4), other were kept to their recommended values (hyper-parameters not listed were set to their default value in scikit-learn).

**Table 4.** Model hyper-parameters and tested values. n is the total number of combinations tested per algorithm.

| Algorithm | Description | Hyper-parameter | Values | n |
|---|---|---|---|---|
| LASSO | Linear regressor that performs variable selection and regularisation. | Regularization parameter alpha | 13 regularly log-spaced points in $[10^{-5}, 1]$ | 13 |
| RF | Ensemble regressor that averages the output of multiple regression trees. | Maximum depth of the tree | 7 values with step of 5 in [10, 40] | 252 |
| | | Maximum number of features at split | All n input features or $\sqrt{n}$ | |
| | | Number of trees | [100, 250, 500] | |

| | | Minimum number of samples required to split an internal node | Fraction of n samples: 6 values with step 0.2 in [0.2, 0.8] | |
|---|---|---|---|---|
| SVR_lin | Regressor that finds the optimal regression hyperplane so that most training samples lie within a certain margin around it. | Gamma | 7 regularly log-spaced points in $[10^{-2}, 10^{2}]$ | 392 |
| | | Epsilon | 7 regularly log-spaced points in $[10^{-6}, 10^{0.5}]$ | |
| | | Regularization parameter C | 8 regularly log-spaced points in $[10^{-5}, 10^{2}]$ | |
| SVR_rbf | Regressor that finds optimal regression hyperplane maps by mapping the input data to a high dimensional feature space using non-linear kernel functions, here, radial basis functions. | Gamma | 7 regularly log-spaced points in $[10^{-2}, 10^{2}]$ | 392 |
| | | Epsilon | 7 regularly log-spaced points in $[10^{-6}, 10^{0.5}]$ | |
| | | Regularization parameter C | 8 regularly log-spaced points in $[10^{-5}, 10^{2}]$ | |
| GBR | Ensemble of shallow trees in sequence where each new tree minimises the residuals of the previous tree. | Learning rate | [0.01, 0.05, 0.1] | 54 |
| | | Maximum depth of the tree | [10, 20, 40] | |
| | | Number of boosting stages | [100, 250 500] | |
| | | Minimum number of samples required to split an internal node | Fraction of n samples: 6 values with step 0.14 in [0.1, 0.8] | |
| MLP | Artificial neural network that uses a nonlinear weighted combination of the features to predict the target variable. | Regularization parameter alpha | 6 regularly log-spaced points in $[10^{-5}, 10^{-1}]$ | 600 |
| | | Activation functions | rectified linear unit and hyperbolic tangent | |
| | | Learning rate | Constant (0.001) and adaptive | |
| | | Hidden layers/units | 2 layers: every pairwise combinations of [16, 32, 48, 64] units | |
| | | | 3 layers: [16, 32, 16], [16, 48, 16], [32, 48, 32], [32, 64, 32], [48, 64, 48], [32, 32, 32], [48, 48, 48], [64, 64, 64], [16, 16, 16] units | |

### 3.2. Benchmark models

The ML models were compared against two benchmark models: the null and the peak NDVI model. The null model is a naïve one in which the yield forecast for a given province-year is the average yield across all other years of that province. The peak NDVI model assumes that yield is linearly correlated with the seasonal maximum of NDVI at the administrative unit (*yield = a × max(NDVI) + b*) and has been used extensively to estimate cereal yield (Becker-Reshef et al., 2010; Franch et al., 2015 and references therein).

Unlike ML models, calibration and performance assessment of benchmark models are achieved with a simple leave-one-full-year out cross validation. In such cross validation, data of one year of the *n* available is held out for testing. The remaining *n*-1 years are used to train the model. Finally, the model is used to predict the yield of the held out year. This procedure is repeated for all the years.

### 3.3. Statistical analysis

*3.3.1.    Model evaluation and selection*

Four prediction accuracy statistics were extracted from the comparison of observed and predicted yields: the root mean square error ($RMSE_p$), the relative $RMSE_p$ percentage ($rRMSE_p$, obtained normalizing by the crop-specific average yield), the mean error ($ME_p$), and the coefficient of determination ($R^2_p$) between modelled and observed yield. These error metrics were computed on yield forecasts at both the provincial and national levels. Provincial level metrics are time average of the metrics computed on the cross-validation folds. Therefore, $R^2_p$ refers to the average coefficient of determination in the spatial dimension. The average $R^2_p$ in the temporal dimension is instead computed by averaging the temporal $R^2$ (computed per province using all years at once) of all provinces. National yield forecasts were computed as a weighted average of the province yields using province average production from official statistics as weighting factor.

In addition, because forecasting performance in poor-yield years are of particular interest for early warning, at the national level we computed the $RMSE_p$ and $rRMSE_p$ for those years belonging to the first quartile of the yield distribution (low-yield years).

All model configurations were trained and tested for each of the eight forecast months. The best model to be used in operations is selected based on average province-level $rRMSE_p$.

*3.3.2.    Practical significance testing*

For each forecast month, we compared the best ML model configuration (*i.e.*, set of input features, feature selection option and ML algorithm) to the benchmark models and to the best configurations of the other ML algorithms.

For the comparison we used Bayesian hypothesis tests (Bayesian correlated t-test in particular; Benavoli et al., 2017) to establish if the mean difference of $rRMSE_p$ between the best ML configuration at each prediction time and the other method was significant. While hypothesis testing in machine learning usually relies on null hypothesis significance tests, these tests suffer from well-known limitations. For instance, statistical significance does not necessarily imply practical significance. With null hypothesis significance testing, enough data can confirm arbitrarily small effects while conceivable differences may fail to yield low *p*-values if there are not sufficient data. Moreover, null hypothesis significance tests cannot verify the null hypothesis and thus cannot recognise equivalent models. But most importantly, null hypothesis significance tests do not answer the question of which hypothesis is the most likely. Bayesian hypothesis tests overcome these limitations by computing the posterior probability of the null (equivalence) and the alternative hypothesis.

It is common to define a Region Of Practical Equivalence (ROPE) to facilitate the interpretation of the posterior probabilities resulting of a Bayesian hypothesis test. The ROPE expresses "the range of parameter values that are equivalent for current practical purposes" (Kruschke and Liddell, 2018). The ROPE thus provides a sensible null hypothesis for the test. The ROPE must be set by users depending on the metric used for the comparison and on domain-specific, and domain-specific (often subjective) definition of practical equivalence. Defining the ROPE as the interval $[-\delta, \delta]$, when comparing two algorithms ($algorithm_1$ and $algorithm_2$), three probabilities can be computed from the posterior:

- *P(algorithm$_1$ < algorithm$_2$)*: the integral of the posterior on the interval $(-\infty, -\delta)$.
- *P(algorithm$_1$ = algorithm$_2$)*: the integral of the posterior over the rope interval $(-\delta, \delta)$.
- *P(algorithm$_1$ > algorithm$_2$)*: the integral of the posterior on the interval $(\delta, \infty)$

Here, practical differences were tested on $rRMSE_p$ by setting a ROPE of 5% and a 90% confidence level (alpha = 0.1). If none of the above probabilities were above a desired confidence level, the outcome of the test was deemed inconclusive. To facilitate visualisation, we colour-coded using the Red-Green-Blue colour space (Fig. 3).

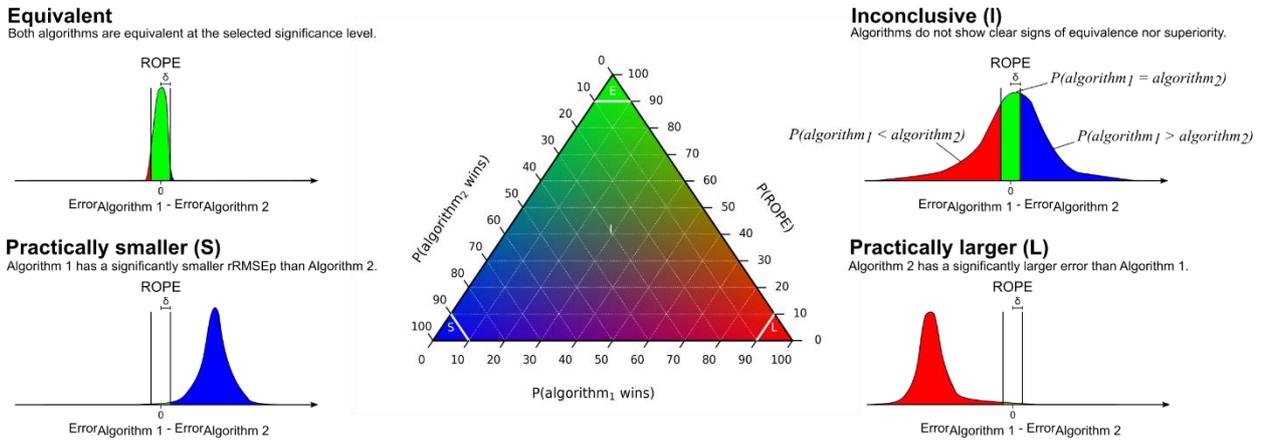

**Fig. 3.** Bayesian hypothesis test and diagnosis to compare two algorithms. The plain white lines delineate the regions of practical significance (at the 90% confidence level). Test outcomes falling outside of these regions are inconclusive.

*3.3.3.      Effect of OHE, feature set and selection*

The effect of feature selection was evaluated by comparing the rRMSE$_p$ obtained with and without each feature selection option. We evaluated these effects at each forecast month by computing the gain in rRMSE$_p$ obtained with different options, algorithm-wise or feature set-wise, to highlight specific impacts.

## 4. Results

### 4.1. Performances by forecasting time

The best ML models were always more accurate than the peak NDVI model regardless of the forecast month (Fig. 4; see also Table 5 for the forecasting period April - July). However, the selection of the most appropriate ML configuration is essential to guarantee this performance improvements as benchmark models outperform a large proportion of the ML models that were trained (Fig. 5). SVR, either with a linear kernel or with a radial basis function kernel, is the most frequently selected algorithm, followed by Lasso and MLP. RF and GBR never achieve the best performances. Adding one-hot-encoded variables about administrative units is always preferred. Remote sensing input features are always selected while meteorological features are used in a minority of cases and always with remote sensing features. Feature selection is frequently activated and the selected feature set contains meteorological variables more often than not.

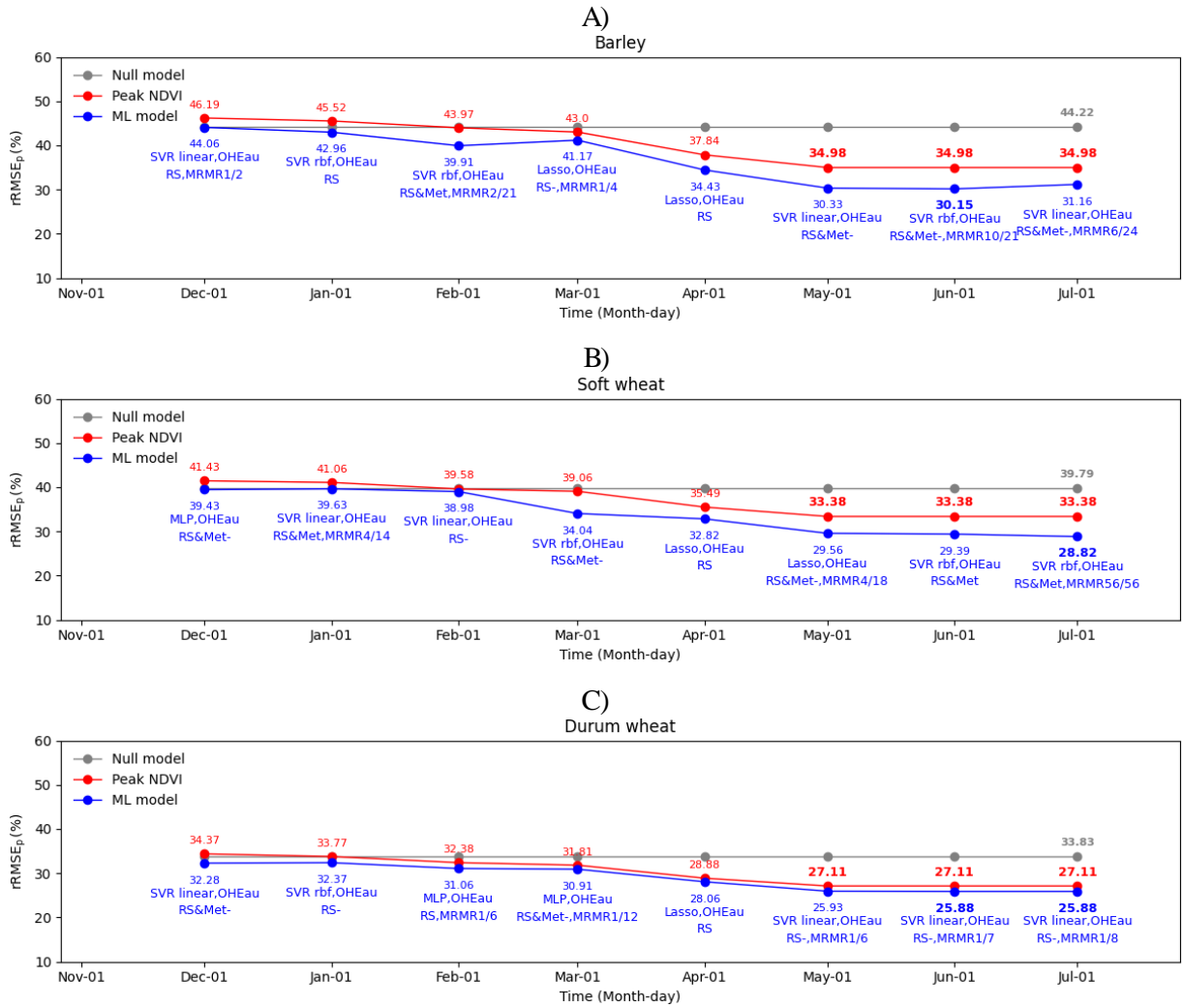

**Fig. 4.** Predictive performances (rRMSE$_p$) of the null model, the peak NDVI model and the best ML model configuration by forecasting time. Numbers in bold highlight to the minimum rRMSE$_p$ achieved. Details of the ML model configuration are given: algorithm, one-hot-encoded variables about administrative unit (OHEau), feature set and mRMR feature selection. When mRMR is activated, the number of retained features over the total is shown. As this fraction can vary in the cross-validation outer loop we report here as an indication the fraction selected when fitting all the data. It is noted that, despite the activation of mRMR feature selection provides the largest accuracy in cross validation, all features may be retained in fitting.

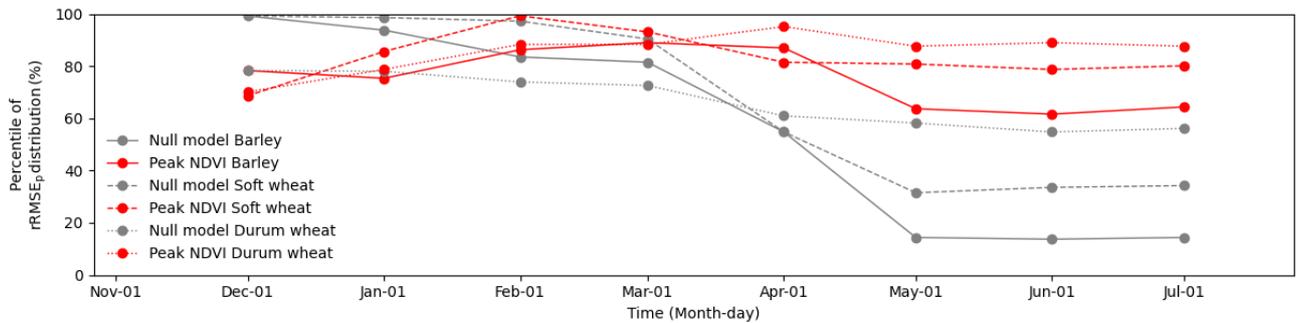

**Fig. 5**. Null and Peak NDVI percentile rank in the rRMSE$_p$ distribution of all tested model configurations.

Taking the May forecasts as example, the selected ML model configuration explained about 50% of the spatial variability ($R^2_p$ equal to 0.47, 0.5 and 0.66 for barley, soft and durum wheat, respectively; Table 5). A slightly lower percentage is explained in the temporal dimension of the province-level yield variability ($R^2_p$ equal to 0.46, 0.42 and 0.34 for barley, soft and durum wheat, respectively).

ML models show greater accuracies at the national level with $R^2_p$ equal to 0.81, 0.74 and 0.61 against 0.52, 0.51 and 0.58 of the peak NDVI benchmark for barley, soft and durum wheat, respectively. When focusing on low-yield years, the forecast accuracy of ML models remained vastly constant, unlike benchmark models. In fact, ML models forecasted yields twice as accurately as the peak NDVI model for barley and soft wheat while similar performances are observed for durum wheat.

**Table 5.** Monthly performance of the null model, the peak NDVI model and the best ML model from April to July. Metrics are computed (1) by held-out cross-validation fold, and then averaged; (2) by province pooling together all years, and then averaged; (3) aggregating yearly province data at national level. Numbers in bold highlight to the highest accuracies.

| Crop | Forecasting time | Model | OHEau | Feat. set | Feat. selection | Selected/Total (fitting) | Average on yearly metrics(1) | | | | Average on province metric(2) | National level aggregation(3) | | | | | | | ΔrRMSE$_p$ (FQ-all years) |
|---|---|---|---|---|---|---|---|---|---|---|---|---|---|---|---|---|---|---|---|
| | | | | | | | | | | | | All years | | | | First quartile | | | |
| | | | | | | | $R^2_p$ | RMSE$_p$ (t/ha) | rRMSE$_p$ (%) | ME$_p$ (t/ha) | $R^2_p$ | $R^2_p$ | RMSE$_p$ (t/ha) | rRMSE$_p$ (%) | ME$_p$ (t/ha) | RMSE$_p$ (t/ha) | rRMSE$_p$ (%) | | |
| Barley | Apr | Lasso | y | RS | n | | **0.24** | **0.42** | **34.43** | 0.00 | **0.31** | **0.48** | **0.26** | **21.74** | -0.01 | **0.32** | **26.58** | 4.84 |
| | | Peak NDVI | - | - | - | | -0.04 | 0.46 | 37.84 | -0.01 | 0.17 | 0.35 | 0.29 | 24.29 | 0.01 | 0.38 | 31.48 | 7.19 |
| | | Null | - | - | - | | -0.46 | 0.54 | 44.22 | **0.00** | -0.13 | -0.13 | 0.39 | 32.01 | **0.00** | 0.51 | 41.93 | 9.92 |
| | May | SVR_linear | y | RS&Met | n | | **0.47** | **0.37** | **30.33** | 0.02 | **0.46** | **0.81** | **0.16** | **13.09** | -0.02 | **0.16** | **13.42** | 0.33 |
| | | Peak NDVI | - | - | - | | 0.09 | 0.42 | 34.98 | -0.01 | 0.27 | 0.52 | 0.25 | 20.78 | 0.00 | 0.33 | 27.28 | 6.50 |
| | | Null | - | - | - | | -0.46 | 0.54 | 44.22 | **0.00** | -0.13 | -0.13 | 0.39 | 32.01 | **0.00** | 0.51 | 41.93 | 9.92 |
| | Jun | SVR_rbf | y | RS&Met | y | 10/21 | **0.47** | **0.36** | **30.15** | 0.00 | **0.47** | **0.76** | **0.18** | **14.66** | -0.004 | **0.21** | **17.45** | 2.79 |
| | | Peak NDVI | - | - | - | | 0.09 | 0.42 | 34.98 | -0.01 | 0.27 | 0.52 | 0.25 | 20.78 | 0.003 | 0.33 | 27.28 | 6.50 |
| | | Null | - | - | - | | -0.46 | 0.54 | 44.22 | **0.00** | -0.13 | -0.13 | 0.39 | 32.01 | **0.000** | 0.51 | 41.93 | 9.92 |
| | Jul | SVR_linear | y | RS&Met | y | 6/24 | **0.44** | **0.38** | **31.16** | 0.01 | **0.42** | **0.70** | **0.20** | **16.60** | -0.01 | **0.24** | **20.16** | 3.56 |
| | | Peak NDVI | - | - | - | | 0.09 | 0.42 | 34.98 | -0.01 | 0.27 | 0.52 | 0.25 | 20.78 | 0.003 | 0.33 | 27.28 | 6.50 |
| | | Null | - | - | - | | -0.46 | 0.54 | 44.22 | 0.00 | -0.13 | -0.13 | 0.39 | 32.01 | **0.000** | 0.51 | 41.93 | 9.92 |
| Soft wheat | Apr | Lasso | y | RS | n | | **0.40** | **0.46** | **32.82** | 0.00 | **0.29** | **0.59** | **0.24** | **16.88** | 0.0004 | **0.22** | **15.47** | -1.42 |
| | | Peak NDVI | - | - | - | | 0.26 | 0.50 | 35.49 | -0.01 | 0.17 | 0.38 | 0.29 | 20.73 | 0.0040 | 0.34 | 24.35 | 3.62 |
| | | Null | - | - | - | | 0.05 | 0.56 | 39.79 | **0.00** | -0.13 | -0.13 | 0.39 | 27.97 | **0.0000** | 0.51 | 36.66 | 8.68 |
| | May | Lasso | y | RS&Met | y | 4/18 | **0.50** | **0.41** | **29.56** | 0.01 | **0.42** | **0.74** | **0.19** | **13.51** | 0.004 | **0.17** | **11.91** | -1.60 |
| | | Peak NDVI | - | - | - | | 0.33 | 0.47 | 33.38 | -0.01 | 0.27 | 0.51 | 0.26 | 18.41 | 0.005 | 0.31 | 22.34 | 3.93 |
| | | Null | - | - | - | | 0.05 | 0.56 | 39.79 | **0.00** | -0.13 | -0.13 | 0.39 | 27.97 | **0.000** | 0.51 | 36.66 | 8.68 |
| | Jun | SVR_rbf | y | RS&Met | n | | **0.55** | **0.41** | **29.39** | 0.02 | **0.46** | **0.84** | **0.15** | **10.62** | -0.002 | **0.16** | **11.11** | 0.49 |
| | | Peak NDVI | - | - | - | | 0.33 | 0.47 | 33.38 | -0.01 | 0.27 | 0.51 | 0.26 | 18.41 | 0.005 | 0.31 | 22.34 | 3.93 |
| | | Null | - | - | - | | 0.05 | 0.56 | 39.79 | **0.00** | -0.13 | -0.13 | 0.39 | 27.97 | **0.000** | 0.51 | 36.66 | 8.68 |
| | Jul | SVR_rbf | y | RS&Met | y | 56/56 | **0.56** | **0.40** | **28.82** | -0.01 | **0.46** | **0.85** | **0.14** | **10.24** | 0.024 | **0.16** | **11.69** | 1.46 |
| | | Peak NDVI | - | - | - | | 0.33 | 0.47 | 33.38 | -0.01 | 0.27 | 0.51 | 0.26 | 18.41 | 0.005 | 0.31 | 22.34 | 3.93 |
| | | Null | - | - | - | | 0.05 | 0.56 | 39.79 | **0.00** | -0.13 | -0.13 | 0.39 | 27.97 | **0.000** | 0.51 | 36.66 | 8.68 |
| Durum wheat | Apr | Lasso | y | RS | n | | **0.60** | **0.41** | **28.06** | -0.01 | **0.23** | **0.45** | **0.24** | **16.87** | 0.01 | **0.24** | **16.60** | -0.27 |
| | | Peak NDVI | - | - | - | | 0.57 | 0.42 | 28.88 | -0.01 | 0.20 | 0.45 | 0.24 | 16.78 | 0.01 | 0.27 | 18.72 | 1.94 |
| | | Null | - | - | - | | 0.42 | 0.49 | 33.83 | **0.00** | -0.13 | -0.13 | 0.35 | 24.13 | **0.00** | 0.48 | 33.24 | 9.12 |
| | May | SVR_linear | y | RS- | y | 1/6 | **0.66** | **0.38** | **25.93** | -0.01 | **0.34** | **0.61** | **0.206** | **14.20** | 0.01 | 0.26 | 18.26 | 4.06 |
| | | Peak NDVI | - | - | - | | 0.63 | 0.39 | 27.11 | 0.00 | 0.30 | 0.58 | 0.214 | 14.77 | 0.01 | **0.25** | **17.44** | 2.67 |
| | | Null | - | - | - | | 0.42 | 0.49 | 33.83 | **0.00** | -0.13 | -0.13 | 0.35 | 24.13 | **0.00** | 0.48 | 33.24 | 9.12 |
| | Jun | SVR_linear | y | RS- | y | 1/7 | **0.67** | **0.38** | **25.88** | -0.01 | **0.34** | **0.61** | **0.205** | **14.14** | 0.01 | 0.26 | 18.10 | 3.96 |
| | | Peak NDVI | - | - | - | | 0.63 | 0.39 | 27.11 | 0.00 | 0.30 | 0.58 | 0.214 | 14.77 | 0.01 | **0.25** | **17.44** | 2.67 |
| | | Null | - | - | - | | 0.42 | 0.49 | 33.83 | **0.00** | -0.13 | -0.13 | 0.35 | 24.13 | **0.00** | 0.48 | 33.24 | 9.12 |
| | Jul | SVR_linear | y | RS- | y | 1/8 | **0.67** | **0.38** | **25.88** | -0.01 | **0.34** | **0.61** | **0.205** | **14.14** | 0.01 | 0.26 | 18.10 | 3.96 |
| | | Peak NDVI | - | - | - | | 0.63 | 0.39 | 27.11 | 0.00 | 0.30 | 0.58 | 0.214 | 14.77 | 0.01 | **0.25** | **17.44** | 2.67 |
| | | Null | - | - | - | | 0.42 | 0.49 | 33.83 | **0.00** | -0.13 | -0.13 | 0.35 | 24.13 | **0.00** | 0.48 | 33.24 | 9.12 |

Despite the best machine learning model configurations delivered the most accurate forecasts for each forecast date and crop, differences between other algorithms or benchmarks were not always practically significant (Fig. 6). Among machine learning models, we report a high variability of practical differences for all three crops. Tests for LASSO, MLP and SVR (both rbf and linear) were commonly inconclusive (I), that is, despite being inferior to the best machine learning model, the probability of observing a difference larger than 5% did not reach the selected confidence level (90%). The peak NDVI benchmark had a significantly lower (L) accuracy across the season for barley, and in the second half of the season for soft wheat and in the first part of the season for durum wheat. The null benchmark displayed more consistency: it performed similarly to the best machine learning method until March (fourth forecast) and then underperformed.

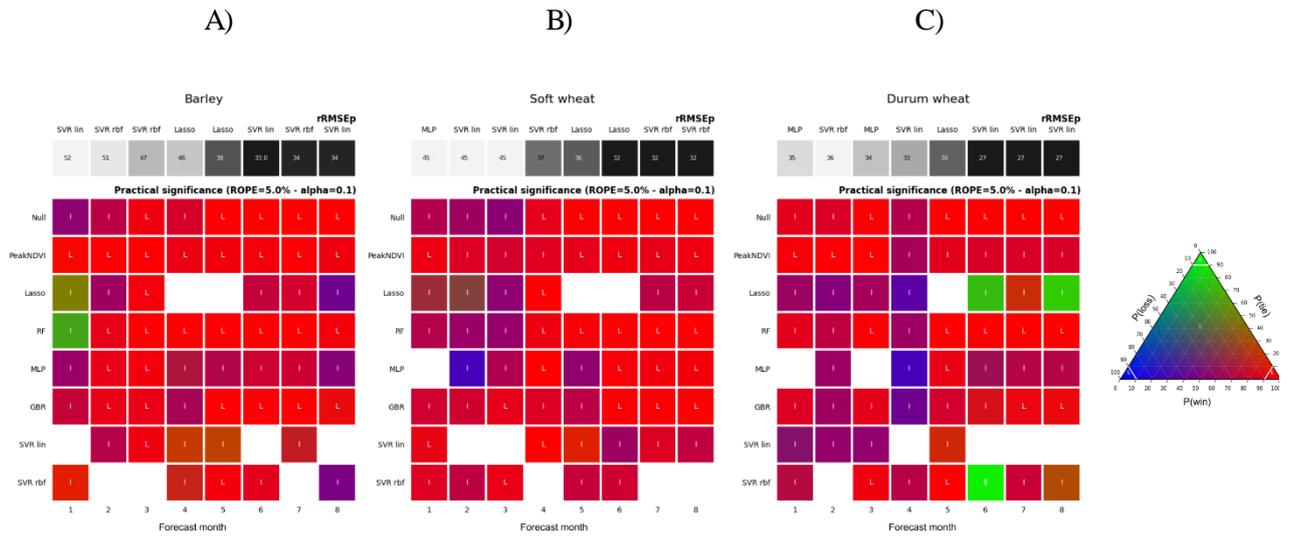

**Fig. 6.** Bayesian hypothesis testing for barley (A), soft (B) and durum wheat (C). The best model configuration at each forecasting time (top row in grey) is tested against benchmark models and all other ML algorithms best configurations (matrix below). The top part of each subfigure indicates the average rRMSE$_p$ of the best algorithm at each forecasting time. The bottom part uses Red-Green-Blue colour space to represents the posterior probabilities for the best algorithm to be better (reddish colours), worse (bluish colours) or equivalent (greenish colours) of the other models for a ROPE of 5% rRMSE$_p$. Differences with suboptimal algorithms are practically larger, equivalent or smaller (L, E, S) if the respective probabilities are greater than 0.9, otherwise the test is inconclusive (I).

### 4.2. Effects of feature sets and feature selection

The use of One-Hot Encoded features representing the administrative units (OHEau) appears instrumental to increase the accuracy of all ML model configurations (Fig. 7A) with a median reduction of rRMSE$_p$ ranging from 13 to 1%. Despite some variability in the improvement between ML algorithms, a general decrease of improvement with increasing forecasting time is observable. Given the utility of OHEau, the following analyses were restricted to model configuration using that option to improve readability of the results. Meteorological features were mainly relevant early in the season (Fig. 7B). From April onward, they did not allow a sensible reduction of the error unlike remote sensing features.

Feature selection helped increase accuracy in 63.5 % of the cases. Gains increased as the season progressed for Lasso, SVR linear and MLP while they surprisingly decreased for RF and GBR (Fig. 7C). They remained constant for SVR rbf. Greatest and most variable benefits were observed for the RS&Met feature set (the largest feature set) and smallest and less variable benefits were observed for the RS and RS- feature sets (Fig. 7D).

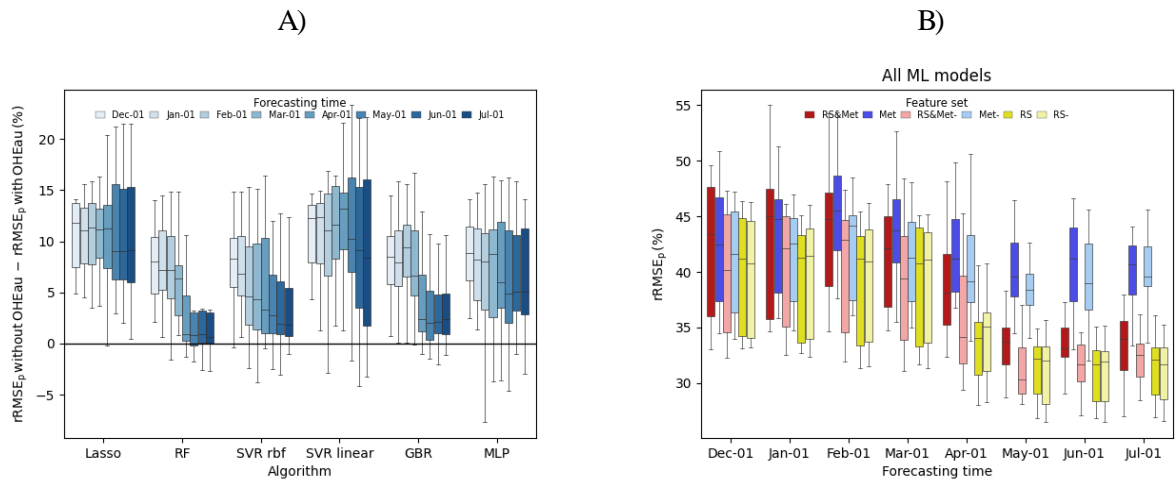

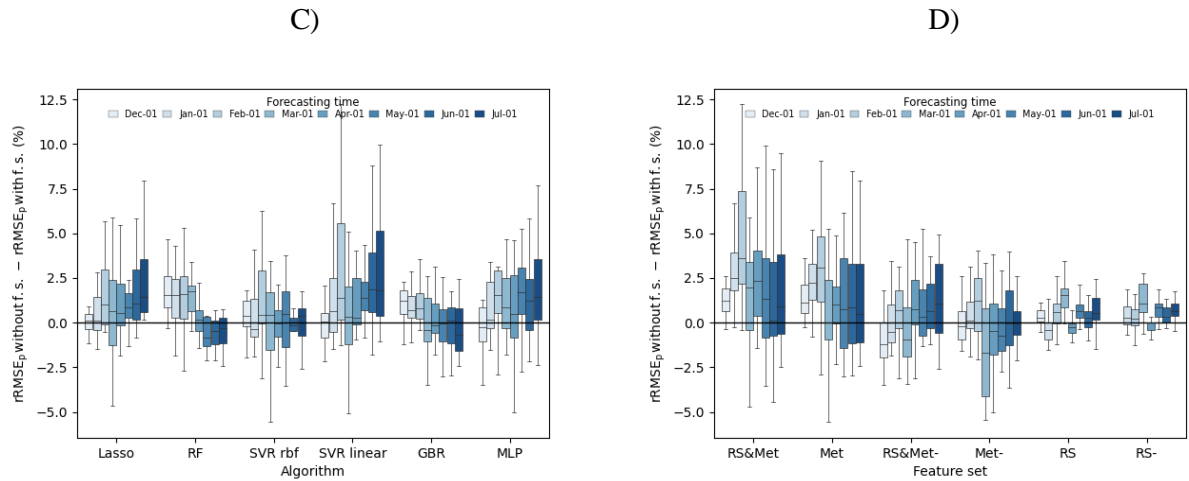

**Fig. 7.** Impact of feature sets and feature selection on forecast accuracy. Statistics computed on all crops, only configurations with OHEau features are shown for graphs in B, C and D. A) Reduction of rRMSE$_p$ due to the use of OHEau by ML algorithm and forecasting time. B) rRMSE$_p$ by forecast time and feature set. C) and D) reduction of rRMSE$_p$ due to feature selection by ML algorithm and feature set. The box represents the sample lower quartile, the median and the upper quartile. Whiskers extend to extend to points that lie within 1.5 interquartile ranges of the lower and upper quartile. Supplementary Fig. S5 reports the same data of A) by input feature data set.

## 5. Discussion

### 5.1. ML model forecasting performances

Selecting the best ML model configuration outperformed the benchmark models at any forecasting time and for any crop (Fig. 4). Yet, there is no silver bullet algorithm and settings such that each of them requires appropriate calibration. In fact, the sole use of a ML algorithm or the inclusion of meteorological variables in the input features is not a sufficient condition for improved performances compared to the peak NDVI benchmark model (Fig. 5).

Early predictions of ML models (*i.e.*, December and January) outperformed the null model by a small margin and thus had little added value (Fig. 4). In February-March (depending on crop type), they started to provide forecasts with improved performance. In April, we observed prediction errors in the magnitude of range of 28-34 % rRMSE$_p$ on province forecasts and 17-21% on national level forecasts. Minimum error was achieved in June for barley and durum wheat and in July for soft wheat. However, accuracy gains were minimal after the significant error reduction observed in early May (and thus before the start of the harvesting operations).

Errors were comparable for barley and soft wheat and were lower for durum wheat. Nonetheless, the gains offered by machine learning as compared to peak NDVI was not as large as for durum wheat as for the other two crops. We think that this could reflect the more extensive cropping in the north of Algeria for durum wheat than for barley and soft wheat (Supplementary Fig. S5). In the North precipitations are more abundant and soils more fertile. The importance of meteorological variables in such area may thus be reduced and the seasonal trajectory of NDVI alone might be already well suited for yield forecasting. Indeed, from April onward, the best ML model configurations for durum wheat also relied only on NDVI. On the contrary, meteorological features were frequently retained (always after April) for barley and soft wheat, for which machine learning deliver higher performance gains. In addition, higher accuracy of the null model (*i.e.*, average yield by province) for durum wheat compared to the other crops indicates lower inter-annual variability (see also Fig. S1) and may further explain the reduced added value of machine learning.

Our study presents a rigorous framework for out-of-sample validation: the test data was not used for model training. Leaving one year out during cross-validation allowed us to evaluate the true forecasting accuracy of our models, which is lacking in more than half of the studies analysed in Schauberger et al. (2020). Therefore, accuracy levels similar to those reported in this paper can be expected in operational forecasting. Forecast accuracy (Table 5) is comparable to that of other studies forecasting cereal yields in the same climatic and geographic area. It is higher than what Meroni et al. (2013) reported for durum wheat yield forecast (0.35 t/ha) in the neighbouring country of Tunisia using cumulative fraction of photosynthetically active radiation, and it is lower than what of Balaghi et al. (2008) reported in Morocco for near harvest forecasts (average province level MAE of 0.21 t/ha against 0.31 of this study) using stepwise regression on NDVI, temperature and precipitation data at the province level.

Performances of ML models appear to be more stable than those of peak NDVI benchmark when considering only the low-yield years, an important property for yield forecasting in the food security and early warning context. The variation of $rRMSE_p$ when focussing on first quartile of the yield distribution (Table 5, $\Delta rRMSE_p$ = $rRMSE_p$ on first quartile – $rRMSE_p$ on all years) is in the majority of cases smaller for ML models as compared to peak NDVI benchmark. Taking the forecasting times of April and May as an example, we observed that with the exception of durum wheat in May, ML models provide a smaller reduction of performances (and for soft wheat an improvement of performances) as compared to peak NDVI when focusing on the poorest yield years.

### 5.2. Effect of province information, feature sets and feature selection

In agreement with Wolanin et al. (2020), we observed that ML models performed better when information on the province was included (Fig. 7A) indicating that there are inter-province differences that were not captured neither by observed NDVI nor by climate drivers. Province-level information is implicitly available to peak NDVI models because these are tuned per province.

At the beginning of the season, the performance of the meteorological-only (Met and Met-) and NDVI-only models (RS and RS-) is comparable (Fig. 7B). As the season progresses, the performance of NDVI-based models improve while meteorological-only models sets do not. This illustrates that information initially provided by meteorological growth drivers is incorporated into NDVI, reflecting actual crop growth (Waldner et al. 2019). It is further noted that in the Algerian setting the inclusion of monthly minimum and maximum 10-day average temperatures (present in set Met) has not benefit. Nevertheless, the use of a reduced set containing both meteorological and remote sensing features (RS&Met-) appears to be useful between May and July when weather may influence first the flowering (e.g. heat stress in April reducing flower fertility and thus, the of grains per spike) and later on the grain weight, by widening or reducing the grain filling period (e.g., mild temperature could extend it while droughts accelerate it).

Overall, feature selection with minimum redundancy maximum relevance boosted model performances. However, exceptions exist (Fig. 4). On the one hand, these were likely due to both data dimensionality and collinearity, which in our analyses depend on the input feature set and forecast month. Indeed, the greatest benefits of feature selection were observed for the largest feature set (*i.e.*, RS&Met). On the other hand, the nature of the algorithm themselves could explain these divergences. LASSO, SVR linear and MLP benefited from feature selection and gains increased later in the season when more features became available. The opposite was observed for tree-based methods (RF and GBR). This suggests that these algorithms can extract relevant feature more efficiently than mRMR and/or are robust to redundant features. Given the observed variability in performance, we recommend that feature selection should be systematically tested (Fig. 7C and D).

Contrary to feature set selection (manual feature engineering) that operates a semantic selection on the type of variables considered (*i.e.*, including or excluding one or more specific variables for the whole period), automatic mRMR feature selection operates on single features (*i.e.*, specific variables at a specific times). This complicates the interpretation of feature selection impact on performance often observable in March in Fig. 7D. Despite the lack of interpretation, this month-specific reduction of feature selection utility reiterates the importance of data-driven test of ML algorithm configurations.

There is often a tension between manually engineering feature sets based on domain knowledge and letting machines selecting them based on data. Our results showed that knowledge-driven feature selection slightly outperformed fully data-driven feature selection (Fig. S6). However, largest improvements can be achieved when both selection methods are combined.

### 5.3. Perspectives

Our workflow was limited to few variables: NDVI, temperature, global radiation and rainfall. Adding more variables will likely boost the forecast accuracy. For instance, other vegetation indices or specific spectral bands play a role to limit soil and atmospheric disturbances (Kouadio et al., 2014) or to avoid NDVI saturation at dense canopy cover (Peralta et al., 2016). Other variables such as vapour pressure deficit, Solar Induce Fluorescence, Ku-band backscatter, thermal-based evapotranspiration, Vegetation Optical Depth and soil moisture from active and passive measurements provide unique information on environmental stresses that improves overall crop yield predictive skill (Cai et al., 2019b; Guan et al., 2017; Martinez-Ferrer et al., 2020).

Increasing the temporal resolution may also capture crop growth more precisely and improve yield estimation as in Waldner et al. (2019). However, some of these data sources have limited archive (<10 years). As yield statistics, access to historical observations is critical to build a sufficiently large training set. Here, 18 years of archive data yielded <410 data points, which is small data for machine learning standards.

One of the specificity of our study is its data-poor context. Not only machine learning models were trained with small data sets but also a static cropland mask was used throughout the period of interest for all three crops. Masks are not a direct input to yield forecasting methods, they are used to filter out irrelevant areas where crops are not grown. Errors in the mask (*e.g.*, the presence of fallows or variations in the cropped area) will thus decrease the signal to noise ratio, which will likely decrease accuracy. As a generic cropland mask was used, crop-specific time series could not be produced so that yield forecasts for the different crops were produced based on the same aggregated time series. Although it may be thus expected that forecasts where a given crops is less prevalent are less accurate because it contributes little to the aggregated time series, we did not find any significant correlation between crop-specific province error and province fraction of total cereal area covered by the crop from official statistics (data not shown).

Mining the MODIS or Landsat archive to produce dynamic cropland masks (*e.g.*, Waldner et al., 2017) is likely to help improve accuracy. Producing crop type maps will be more challenging as crop-specific training data from past years are unavailable.

Finally, it is worth mentioning that the automatic testing pipeline comes at the expenses of high computing resources. Run time for the Algerian case study amounts to 3.3 years of a 4-CPUs machine (20 days on the computer cluster we used). Nevertheless, once model configuration selection is achieved, the resources need for operational yield forecasting are negligible. Future developments will include the integration of the pipeline into the ASAP online system to facilitate yield forecasting to interested practitioners that may upload their country statistics and get the resource demanding tuning process made remotely.

## 6. Conclusion

We introduced a generic and robust machine learning workflow to forecast crop yields with small, public and easily-accessible climate and satellite time series. Our workflow is fully automated and identifies the best model configuration for prediction during the growing season. A rigorous testing procedure is applied to provide historical predictive performances at provincial and national level, also focusing on errors relevant in the food security context, *i.e.*, made when estimating low-yield years. We deployed our workflow in Algeria, where we predicted barley, soft and durum wheat yields on a monthly basis for the 2002 to the 2018 growing seasons. While the best machine learning model always outperformed simple benchmarks, we found that no single model nor feature set combination consistently delivered the best forecasts. Besides, benchmark models yielded more accurate forecasts than a large proportions of the models that were tested. These elements act as a reminder that, while the popularity of machine learning is not underserved, extensive model calibration is required to deliver significant gains in predictive power—the smallness of the data exacerbates this requirement.

**Acknowledgements**

This research was supported by the Joint Research Centre award "A. Royer - Science and Policy for Sustainable Development in Africa".

**References**

Atzberger, C., Vuolo, F., Klisch, A., Rembold, F., Meroni, M., Marcio Pupin, M., Formaggio, A., 2016. Agriculture, in: Thenkabail, P.S. (Ed.), Remote Sensing Handbook. CRC Press, pp. 71–103.
Balaghi, R., Tychon, B., Eerens, H., Jlibene, M., 2008. Empirical regression models using NDVI, rainfall and temperature data for the early prediction of wheat grain yields in Morocco. Int. J. Appl. Earth Obs. Geoinf. 10, 438–452. https://doi.org/10.1016/j.jag.2006.12.001
Becker-Reshef, I., Justice, C., Barker, B., Humber, M., Rembold, F., Bonifacio, R., Zappacosta, M., Budde, M., Magadzire, T., Shitote, C., Pound, J., Constantino, A., Nakalembe, C., Mwangi, K., Sobue, S., Newby, T., Whitcraft, A., Jarvis, I., Verdin, J., 2020. Strengthening agricultural decisions in countries at risk of food insecurity: The GEOGLAM Crop Monitor for Early Warning. Remote Sens. Environ. 237, 111553. https://doi.org/10.1016/j.rse.2019.111553
Becker-Reshef, I., Vermote, E., Lindeman, M., Justice, C., 2010. A generalized regression-based model for forecasting winter wheat yields in Kansas and Ukraine using MODIS data. Remote Sens. Environ. 114, 1312–1323. https://doi.org/10.1016/j.rse.2010.01.010

Benavoli, A., Corani, G., Demšar, J., Zaffalon, M., 2017. Time for a change: A tutorial for comparing multiple classifiers through Bayesian analysis. J. Mach. Learn. Res. 18, 1–36.

Benmehaia, A.M., Merniz, N., Oulmane, A., 2020. Spatiotemporal analysis of rainfed cereal yields across the eastern high plateaus of Algeria: an exploratory investigation of the effects of weather factors. Euro-Mediterranean J. Environ. Integr. 5, 1–12. https://doi.org/10.1007/s41207-020-00191-x

Breiman, L., 2001. Random Forests. Mach. Learn. 45, 5–32.

Cai, Y., Guan, K., Lobell, D., Potgieter, A.B., Wang, S., Peng, J., Xu, T., Asseng, S., Zhang, Y., You, L., Peng, B., 2019a. Integrating satellite and climate data to predict wheat yield in Australia using machine learning approaches. Agric. For. Meteorol. 274, 144–159. https://doi.org/10.1016/j.agrformet.2019.03.010

Cai, Y., Guan, K., Lobell, D., Potgieter, A.B., Wang, S., Peng, J., Xu, T., Asseng, S., Zhang, Y., You, L., Peng, B., 2019b. Integrating satellite and climate data to predict wheat yield in Australia using machine learning approaches. Agric. For. Meteorol. 274, 144–159. https://doi.org/10.1016/j.agrformet.2019.03.010

FAO, IFAD, UNICEF, WFP, WHO, 2018. The State of Food Security and Nutrition in the World 2018. Building climate resilience for food security and nutrition. FAO, Rome.

Franch, B., Vermote, E.F., Becker-Reshef, I., Claverie, M., Huang, J., Zhang, J., Justice, C., Sobrino, J.A., 2015. Improving the timeliness of winter wheat production forecast in the United States of America, Ukraine and China using MODIS data and NCAR Growing Degree Day information. Remote Sens. Environ. 161, 131–148. https://doi.org/10.1016/j.rse.2015.02.014

Friedman, J.H., 2001. Greedy function approximation: A gradient boosting machine. Ann. Stat. 29, 1189–1232. https://doi.org/10.1214/aos/1013203451

Fritz, S., See, L., Bayas, J.C.L., Waldner, F., Jacques, D., Becker-Reshef, I., Whitcraft, A., Baruth, B., Bonifacio, R., Crutchfield, J., Rembold, F., Rojas, O., Schucknecht, A., Van der Velde, M., Verdin, J., Wu, B., Yan, N., You, L., Gilliams, S., Mücher, S., Tetrault, R., Moorthy, I., McCallum, I., 2019. A comparison of global agricultural monitoring systems and current gaps. Agric. Syst. 168, 258–272. https://doi.org/10.1016/j.agsy.2018.05.010

Funk, C., Peterson, P., Landsfeld, M., Pedreros, D., Verdin, J., Shukla, S., Husak, G., Rowland, J., Harrison, L., Hoell, A., Michaelsen, J., 2015. The climate hazards infrared precipitation with stations—a new environmental record for monitoring extremes. Sci. Data 2, 150066. https://doi.org/10.1038/sdata.2015.66

Goodfellow, I., Bengio, Y., Courville, A., 2016. Deep Learning Book, Deep Learning. MIT Press, Cambridge, MA.

Guan, K., Wu, J., Kimball, J.S., Anderson, M.C., Frolking, S., Li, B., Hain, C.R., Lobell, D.B., 2017. The shared and unique values of optical, fluorescence, thermal and microwave satellite data for estimating large-scale crop yields. Remote Sens. Environ. 199, 333–349. https://doi.org/10.1016/j.rse.2017.06.043

Johnson, M.D., Hsieh, W.W., Cannon, A.J., Davidson, A., 2016. Crop yield forecasting on the Canadian Prairies by remotely sensed vegetation indices and machine learning methods. Agric. For. Meteorol. 219, 74–84. https://doi.org/10.1016/j.agrformet.2015.11.003

Kruschke, J.K., Liddell, T.M., 2018. The Bayesian New Statistics: Hypothesis testing, estimation, meta-analysis, and power analysis from a Bayesian perspective. Psychon. Bull. Rev. 25, 178–206. https://doi.org/10.3758/s13423-016-1221-4

López-lozano, R., Duveiller, G., Seguini, L., Meroni, M., García-condado, S., Hooker, J., Leo, O., Baruth, B., 2015. Agricultural and Forest Meteorology Towards regional grain yield forecasting with 1 km-resolution EO biophysical products : Strengths and limitations at pan-European level. Agric. For. Meteorol. 206, 12–32. https://doi.org/10.1016/j.agrformet.2015.02.021

Martinez-Ferrer, L., Piles, M., Camps-Valls, G., 2020. Crop Yield Estimation and Interpretability With Gaussian Processes. IEEE Geosci. Remote Sens. Lett. 1–5. https://doi.org/10.1109/lgrs.2020.3016140

Mateo-Sanchis, A., Piles, M., Muñoz-Marí, J., Adsuara, J.E., Pérez-Suay, A., Camps-Valls, G., 2019. Synergistic integration of optical and microwave satellite data for crop yield estimation. Remote Sens. Environ. 234, 111460. https://doi.org/10.1016/j.rse.2019.111460

Meroni, M., Fasbender, D., Rembold, F., Atzberger, C., Klisch, A., 2019a. Near real-time vegetation anomaly detection with MODIS NDVI : Timeliness vs . accuracy and effect of anomaly computation options. Remote Sens. Environ. 221, 508–521. https://doi.org/10.1016/j.rse.2018.11.041

Meroni, M., Marinho, E., Sghaier, N., Verstrate, M.M., Leo, O., 2013. Remote sensing based yield estimation in a stochastic framework - Case study of durum wheat in Tunisia. Remote Sens. 5. https://doi.org/10.3390/rs5020539

Meroni, M., Rembold, F., Urbano, F., Csak, G., Lemoine, G., Kerdiles, H., 2019b. The warning classification

scheme of ASAP – Anomaly hot Spots of Agricultural Production, v4.0, JRC Technical Report. https://doi.org/10.2760/798528

Meroni, M., Verstraete, M.M., Rembold, F., Urbano, F., Kayitakire, F., 2014. A phenology-based method to derive biomass production anomalies for food security monitoring in the Horn of Africa. Int. J. Remote Sens. 35. https://doi.org/10.1080/01431161.2014.883090

Pedregosa, F., Varoquaux, G., Gramfort, A., Michel, V., Thirion, B., Grisel, O., Blondel, M., Prettenhofer, P., Weiss, R., Dubourg, V., Vanderplas, J., Passos, A., Cournapeau, D., Brucher, M., Perrot, M., Duchesnay, E., 2011. Scikit-learn: Machine Learning in Python. J. Mach. Learn. Res. 12, 2825–2830.

Peng, H., Long, F., Ding, C., 2005. Feature Selection Based on Mutual Information: Criteria of Max-Dependency, Max-Relevance, and Min-Redundancy 27, 1226–1238. https://doi.org/10.1109/TPAMI.2005.159

Pérez-Hoyos, A., Udías, A., Rembold, F., 2020. Integrating multiple land cover maps through a multi-criteria analysis to improve agricultural monitoring in Africa. Int. J. Appl. Earth Obs. Geoinf. 88, 102064. https://doi.org/10.1016/j.jag.2020.102064

Rembold, F., Atzberger, C., Savin, I., Rojas, O., 2013. Using Low Resolution Satellite Imagery for Yield Prediction and Yield Anomaly Detection. Remote Sens. 5, 1704–1733. https://doi.org/10.3390/rs5041704

Rembold, F., Meroni, M., Urbano, F., Csak, G., Kerdiles, H., Perez-Hoyos, A., Lemoine, G., Leo, O., Negre, T., 2018. ASAP: a new global early warning system to detect Anomaly hot Spots of Agricultural Production for food security analysis. Agric. Syst. In press. https://doi.org/https://doi.org/10.1016/j.agsy.2018.07.002

Rosenblatt, F., 1961. Principles of Neurodynamics: Perceptrons and the Theory of Brain Mechanisms. Spartan, New York.

Rouse, J.W., Haas, R.H., Schell, J.A., Deering, D.W., Harlan, J.C., 1974. Monitoring the vernal advancements and retro gradation of natural vegetation. Greenbelt, MD.

Schauberger, B., Jägermeyr, J., Gornott, C., 2020. A systematic review of local to regional yield forecasting approaches and frequently used data resources. Eur. J. Agron. 120, 126153. https://doi.org/10.1016/j.eja.2020.126153

Tibshirani, R., 1996. Regression Shrinkage and Selection via the Lasso. J. R. Stat. Soc. Ser. B 58, 267–288.

Vapnik, V., Golowich, S.E., Smola, A., 1997. Support vector method for function approximation, regression estimation, and signal processing. Adv. Neural Inf. Process. Syst. 281–287.

Waldner, F., Hansen, M.C., Potapov, P. V., Löw, F., Newby, T., Ferreira, S., Defourny, P., 2017. National-scale cropland mapping based on spectral-temporal features and outdated land cover information. PLoS One 12, 1–24. https://doi.org/10.1371/journal.pone.0181911

Waldner, F., Horan, H., Chen, Y., Hochman, Z., 2019. High temporal resolution of leaf area data improves empirical estimation of grain yield. Sci. Rep. 9, 1–14. https://doi.org/10.1038/s41598-019-51715-7

Wolanin, A., Mateo-Garciá, G., Camps-Valls, G., Gómez-Chova, L., Meroni, M., Duveiller, G., Liangzhi, Y., Guanter, L., 2020. Estimating and understanding crop yields with explainable deep learning in the Indian Wheat Belt. Environ. Res. Lett. 15. https://doi.org/10.1088/1748-9326/ab68ac

Zhang, L., Zhang, Z., Luo, Y., Cao, J., Tao, F., 2020. Combining optical, fluorescence, thermal satellite, and environmental data to predict county-level maize yield in China using machine learning approaches. Remote Sens. 12. https://doi.org/10.3390/RS12010021

Zhang, X., Friedl, M.A., Schaaf, C.B., Strahler, A.H., Hodges, J.C.F., Gao, F., Reed, B.C., Huete, A., 2003. Monitoring vegetation phenology using MODIS 84, 471–475.